\documentclass[preprint,proceedings]{rmaa}

\usepackage{paralist}

\usepackage{psfrag,color}

\SetYear{2003}
\SetConfTitle{IAU Coll. 194: Interacting Binaries in the Galaxy and Beyond}

\title{Far-Ultraviolet Surveys of Globular Clusters: Searching for the 
  Products of Stellar Collisions and Near Misses} 

\author{
C. Knigge, \altaffilmark{1} 
D.R. Zurek, \altaffilmark{2}
M.M. Shara, \altaffilmark{2}
K.S. Long, \altaffilmark{3}
R.L. Gilliland, and \altaffilmark{3}
P.A. Charles \altaffilmark{1}
}

\altaffiltext{1}{University of Southampton, UK.}
\altaffiltext{2}{American Museum of Natural History, USA.}
\altaffiltext{3}{Space Telescope Science Institute, USA.}

\shortauthor{Knigge et al.}
\shorttitle{FUV Surveys of Globular Clusters}

\fulladdresses{
\item P.A. Charles and C. Knigge: School of Physics \& Astronomy,
  University of Southampton, Southampton SO17 1BJ, UK
  (\email{pac@astro.soton.ac.uk, christian@astro.soton.ac.uk}).
\item R.L. Gilliland and K.S. Long: Space Telescope Science Institute,
  3700 San Martin Drive, Baltimore, MD 21218
(\email{gillil@stsci.edu, long@stsci.edu}).
\item M.M. Shara and D.R. Zurek: American Museum of Natural History,
  79th Street and Central Park West, New York, NY 10024
  (\email{mshara@amnh.org, dzurek@amnh.org}).

}

\listofauthors{W. J. Henney, A. Collaborator, \& L. Author}
\indexauthor{Knigge, C.}
\indexauthor{Zurek, D.R.}
\indexauthor{Shara, M.M.}
\indexauthor{Long, K.S.}
\indexauthor{Gilliland, R.L.}
\indexauthor{Charles, P.A.}

\abstract{

Far-ultraviolet (FUV) observations with the Hubble Space Telescope are
an excellent way to find and study the hot, blue stellar populations in 
the cores of globular clusters. These populations include
dynamically-formed blue stragglers and interacting binaries (such as
cataclysmic variables), i.e. the products of stellar collisions and
near misses. Using the cluster 47 Tuc as an example, we show how the
combination of FUV imaging and slitless spectroscopy can be used to
uncover and study these populations.}

\addkeyword{Stars: Blue Stragglers}
\addkeyword{Globular Clusters: Individual (47 Tucanae)}
\addkeyword{Stars: Novae, Cataclysmic Variables}
\addkeyword{Ultraviolet: General}
\addkeyword{Stars: White Dwarfs}

\begin{document}
\maketitle

\section{Introduction}
\label{sec:intro}

Globular clusters (GCs) are fantastic stellar crash test
laboratories. Violent encounters between binaries and single stars in
dense cluster cores give rise to exotic stellar populations, such as
blue stragglers (BSs), cataclysmic variables (CVs) and low-mass X-ray
binaries. All of these objects have considerably bluer spectral energy
distributions than ``normal'' cluster members. CVs and young white
dwarfs (WDs), in particular, radiate much of their luminosity in the
far-UV. Observations in this waveband  are therefore well suited to
the detection and  characterization of these important stellar
species. We have therefore embarked on a program to study the cores of
GCs at FUV wavelengths, using both imaging and slitless spectroscopy
with STIS and ACS onboard HST. Here, we present results from
observations of our first target, 47 Tucanae (for  details, see Knigge
et al. 2003; 2003) 

\section{Results}

Figure~1 shows a comparison of FUV (STIS/F25QTZ; a bandpass centered
around 1600~\AA) and U-band (WFPC2/F336W) images of the cluster
core. The tremendous difference in crowding between these images
nicely illustrates the ease with which hot stellar populations can be 
detected in the FUV. The color-magnitude diagram produced from this data
is shown in Figure~2. 

As discussed in Knigge et al. (2002), the FUV imaging data has already
allowed us to: (i) find FUV counterparts to {\em all} 
of the Chandra CV candidates in the core of this cluster; (ii) confirm
the CV nature of these candidates based on 
variability and location in the far-UV  
colour-magnitude diagram; (iii) suggest several additional CV
candidates; (iv) detect a clean, well-populated far-UV BSS sequence;
(v) detect numerous young WDs on the upper end of the cooling
curve.

Since crowding is not a serious problem in the FUV, it is possible to
carry out multi-object slitless spectroscopy even in the dense core of
a globular cluster like 47 Tuc. This is an extremely efficient way of
spectroscopically confirming interesting sources, such as CV
candidates. The power of this approach is demonstrated in Figure~3, which
shows the slitless 2-D spectral image of 47~Tuc and the
1-D extracted spectrum of the brightest source, the eclipsing binary
AKO~9. The strong C~{\sc iv} and He~{\sc ii} emission lines in
immediately suggest that AKO~9 is a cataclysmic variable (see
Knigge et al. 2003).

\begin{figure}[t!]\centering
  \includegraphics[width=\columnwidth]{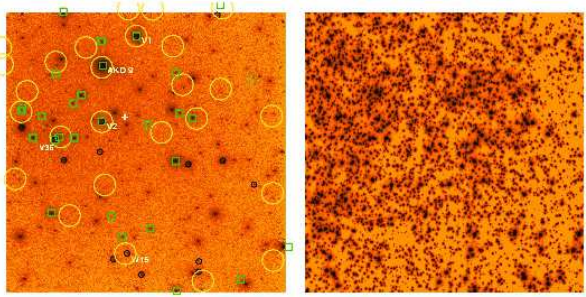}
  \caption{{\em Left Panel:} A  25\arcsec$\times$25\arcsec far-UV
    image of the core of 47~Tuc. The cluster center is marked as a
    cross. The positions of previously known blue objects (squares),
    Chandra x-ray sources (big circles) and CV candidates (small 
    circles) are marked. The four confirmed CVs within the FUV field
    are labelled with their names. {\em Right
    Panel:} The WFPC2/F336W image of the same field. The difference in
    crowding is obvious.}
  \label{fig1}
\end{figure}

\begin{figure}[h!]\centering
  \includegraphics[width=8.0cm]{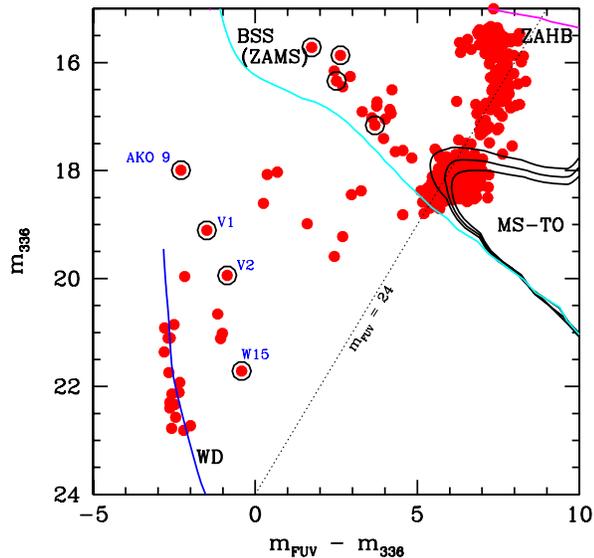}
  \caption{The FUV/optical color-magnitude diagram derived from the
    data in Fig. 1. Variable FUV sources are marked with black
    circles, and the four previously known or suspected CVs are labeled
    with their names. The diagonal line marks the
    FUV completeness limit. The other lines in the diagram indicate
    the expected locations of various stellar
    populations. Unmarked sources between the WD and main sequences
    are potential new CV candidates.} 
\end{figure}

\pagebreak

\adjustfinalcols

\section{Conclusions \& The Future}

Far-ultraviolet surveys are an excellent tool for studying the most
interesting hot stellar populations in globular clusters, including
the dynamically-formed interacting binaries. Here, we have presented
a snapshot of our work in this area, but there is much more to
come. For example, we already have preliminary spectroscopic
confirmation of two additional CV candidates in 47 Tuc, FUV
observations of several other clusters are being analysed, and
additional observations are already scheduled. Watch this space!

\begin{figure}[t!]\centering
  \includegraphics[width=5cm]{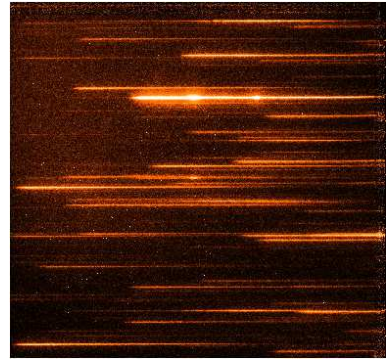}
  \includegraphics[angle=-90, width=5.5cm]{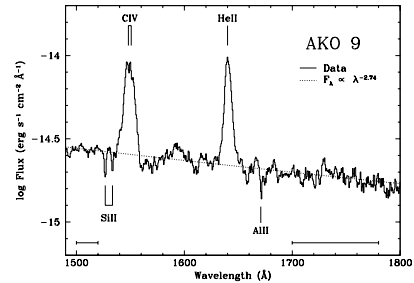}
  \caption{{\em Top Panel:} The 2-D spectral image obtained from our slitless
  FUV spectroscopy of the field shown in Fig. 1. The brightest
  source is the cataclysmic variable AKO~9, which shows clear evidence
  for line emission.{\em Bottom Panel:} The extracted 1-D FUV spectrum
  of AKO~9 (data affected by eclipses was excluded).}
\end{figure}

\end{document}